\documentclass[a4paper,fleqn]{cas-dc}

\usepackage{color}
\usepackage{xspace}
\usepackage{pdfwidgets}
\usepackage{enumerate}

\usepackage[numbers]{natbib}
\usepackage{lineno}
\usepackage{xurl}
\usepackage{latexsym}
\usepackage{graphics}
\usepackage{graphicx}
\usepackage{amsmath}
\usepackage{float}
\usepackage{subfigure}

\def\tsc#1{\csdef{#1}{\textsc{\lowercase{#1}}\xspace}}
\tsc{WGM}
\tsc{QE}

\begin{document}
\let\WriteBookmarks\relax
\def\floatpagepagefraction{1}
\def\textpagefraction{.001}

\shorttitle{Boron-doped Diamond Undulator, Version November 09, 2024}

\shortauthors{H. Backe}

\title [mode = title]{{Planar channeling of 855 MeV electrons in a boron-doped (110) diamond undulator - a case study}}


%

\author[1]{H. Backe}[]
\ead{backe@uni-mainz.de}
\cormark[1]
\author[1]{W. Lauth}[]
\author[1]{P. Klag}[]
\author[2]{Thu Nhi Tran Caliste}[]






\affiliation[1]{organization={{Institute for Nuclear Physics of Johannes Gutenberg-University}},
            addressline={Johann-Joachim-Becher-Weg 45},
            city={Mainz},
            postcode={D-55128},
            state={Germany},
            country={}}

\affiliation[2]{organization={{European Synchrotron Radiation Facility}},
            addressline={71 Avenue des Martyrs},
            city={Grenoble},
            postcode={F-38000},
            state={France},
            country={}}

%
%
%
%
%
%
%


\begin{abstract}
A 4-period diamond undulator with a thickness of 20 $\mu$m was produced with the method of Chemical Vapour Deposition (CVD), applying boron doping, on a straight diamond crystal with an effective thickness of 165.5 $\mu$m. A planar (110) channeling experiment was performed with the 855 MeV electron beam of the Mainz Microtron MAMI accelerator facility to observe the expected undulator peak. The search was guided by simulation calculations on a personal computer. The code is based on the continuum potential picture, and a classical electrodynamic expression which involves explicitly the acceleration of the particle. As a result, an unexpected optimal observation angle was figured out, for which the undulator peak is strongest and the channeling radiation from the backing crystal being significantly suppressed. However, an undulator peak was not observed. Implications for the prepared undulator structure are discussed.

Scatter distributions were measured for a 75 $\mu$m Kapton, a 25 $\mu$m aluminum foils, and a 70.7 $\mu$m diamond plate in random orientation. The results were compared with Molière's scatter theory for amorphous medii. Very good agreement was found for Kapton and aluminum while for diamond the experimental width is 21\% smaller. This reduction is interpreted as coherent scattering suppression in single crystals. At tilted injection of the beam with respect to the (110) plane a clear asymmetry was observed which resembles partial beam deflection. We interpret this phenomenon heuristically as re-channeling.
\end{abstract}





\maketitle

\section{Introduction} \label{intro}
\begin{figure}[tb]
\centering
    \includegraphics[angle=0,scale=0.4,clip]{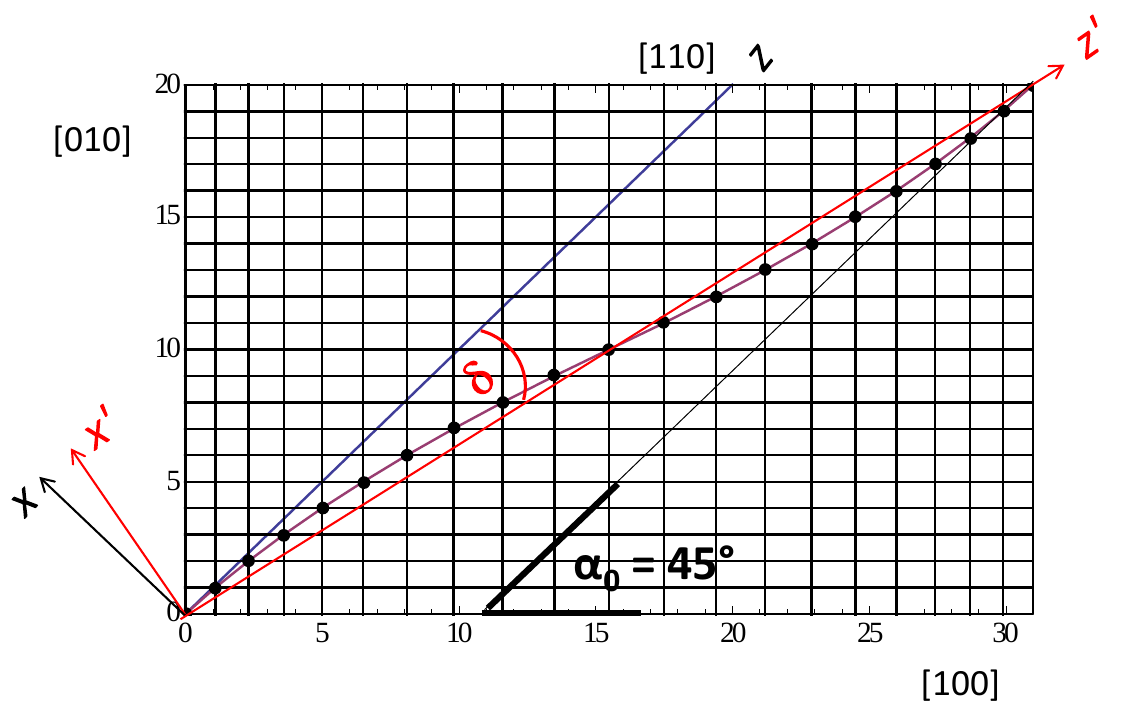}
\caption[]{One period of the undulator crystal. The $z$ direction coincides with the [110] direction of the host crystal, which is not shown but can be imagined as continuation to the right hand border, and makes an angle $\alpha_0$ = 45° with respect to the [100] direction. At observation along the $z^\prime$ axis which makes an angle $\delta$ with respect to the [110] direction an undulator structure is clearly recognizable. Planes are formed perpendicular to the shown plane of drawing, the $y$ - direction.  }\label{UndulatorCrystal}
\end{figure}

There exists a long-lasting considerable interest in the channeling process of ultra-relativistic electrons and positrons at planes of single crystals. Of particular interest is the emission of undulator-like radiation in periodical bent crystals aiming in the construction of compact radiation sources in the MeV range and beyond, see e.g. Korol and Solov'yov \cite{KorS22}. However, it is an experimental challenge to produce crystalline undulators with period lengthes in the order of a few $\mu$m range which are favourable at accelerator facilities delivering electron beams below an energy of about a GeV. Of potential interest are undulators based on  diamond, because of its radiation hardness, which can be produced by Chemical Vapour Deposition (CVD) with periodically varying concentration of boron. The effect is based on the dependence of the lattice constant as function of the boron density resulting in oscillating (110) planes, see Fig. \ref{UndulatorCrystal}. In contrast to Ge-doped silicon undulators, the undulator structure cannot be separated from a host diamond crystal, meaning that in an experiment one is always faced with radiation from the backing.

In this paper we describe experiments with an 4-period diamond undulator with a thickness of 20 $\mu$m at planar (110) channeling, performed with the high quality 855 MeV electron beam of the Mainz Microtron MAMI accelerator facility. The undulator chip was produced with the method of CVD, applying boron doping, on a straight diamond crystal with an effective thickness of 117 $\sqrt{2}$ $\mu$m = 165.5 $\mu$m. During the course of the experiment it became apparent that for an on line optimization a computer code would be desirable which delivers useful results with moderate CPU times and runs, very important, on a personal computer. In contrast to very precise results obtained with the rather sophisticated MBN explorer code \cite{SusB13} employed by Pavlov et al. also for a diamond undulator \cite{PavK21}, for our code a compromise must be found between efficiency and accuracy. 

The paper is organized as follows. In section \ref{undulator} the design parameters of the undulator are described, and in section \ref{Experiment} the experimental setup at MAMI including the measurements. Since the expected undulator peak was not observed, simulation calculations were performed in order to contribute to an explanation of this result. The calculations are based on the continuum potential picture introduced originally by Lindhard \cite{Lin65}. The formalism described in section \ref{basics} and Appendix \ref{appendix A} is based on the papers of one of the authors of this work (H.B.) \cite{Bac22A,Bac24}, extended to undulator crystals, and to the calculation of radiation spectra. For the latter a formula of J.D. Jackson's textbook "Classical Electrodynamics" was used which involves explicitly the acceleration of the particle. It is also the basis of our approach on which an efficient computer code was developed as an aid for an on-line optimization during the course of an experiment. In section \ref{calculations} results are presented and discussed.
The paper includes in section \ref{beamDeflection} experiments and a discussion about the asymmetric scatter distribution at oblique incidence on a straight crystal, and the possibility to deflect by this means parts of the electron beam. An observed reduction of the width of the scatter distribution will be discussed in terms of coherent scattering suppression \cite{MazS20}.  In the conclusions \ref{Conclusions} a possible improved diagnostic tool with the 530 MeV MAMI positron beam \cite{BacL22} is proposed.

\section{The diamond undulator} \label{undulator}
A boron-doped undulator has been produced by Chemical Vapour Deposition with the following design parameters: period length in [100] direction $\lambda_U$ = 5/$\sqrt{2}~\mu$m, number of periods $N_U$ = 4, and a thickness of the host crystal of 117 $\mu$m. The design profile of the boron doping in [100] direction is of triangular shape with the minimum and maximum boron content $n_{B,min}$ = 6 $\cdot 10^{20}$/cm³ and $n_{B,max}$ = 24 $ \cdot10^{20}$/cm³, corresponding to a concentration of $c_{min}$ = 0.00340 and $c_{max}$ = 0.0136, respectively. Analytically, the boron content reads as function of the $z$ coordinate for the first half period $c(z) = c_{min}+2(c_{max}-c_{min})\cdot z/\lambda_U$, $0\leq z\leq \lambda_U/2$. The continuation to $\lambda_U/2< z\leq \lambda_U$ can be obtained by symmetry considerations.

\begin{figure}[tb]
\centering
    \includegraphics[angle=0,scale=0.3,clip]{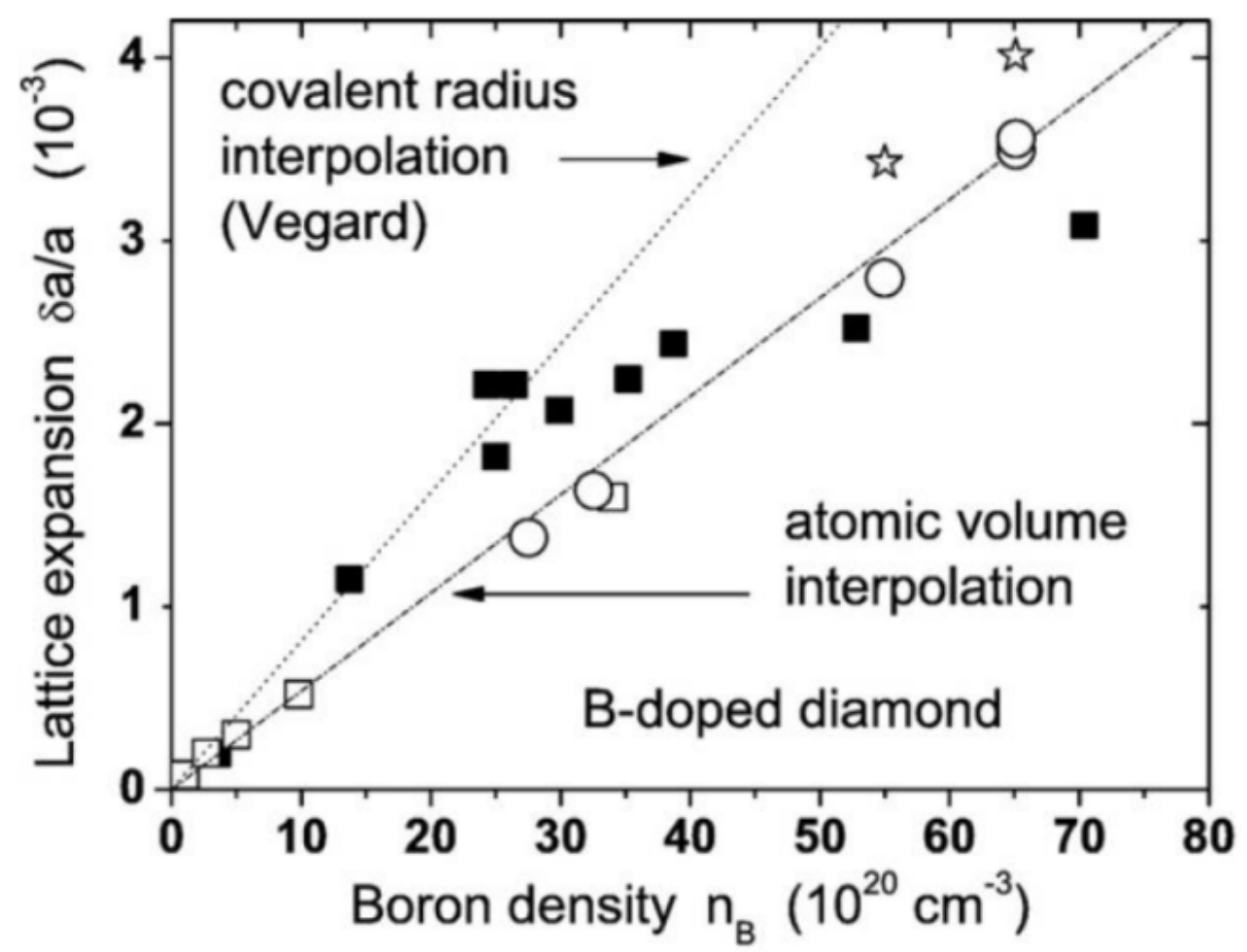}
\caption[]{Lattice expansion as function of the boron density \cite{Ach08}.} \label{BoronCharacteristics}
\end{figure}

With the linear "atomic volume interpolation" function of Fig. \ref{BoronCharacteristics} for $\delta a/a = k~n_e$ with $k = 5.374 \cdot 10^{-25}/\mbox{cm}^{3}$ and $\delta a_{\bot CB}/a_{\bot CB}(z) = k \cdot c(z)$ one obtains the differential equation $dx/dz = -k \cdot c(z)$ for the nominal bent (110) channel. Notice that beside the boundary condition $x(z\rightarrow0) = 0$ also $dx/dz(z\rightarrow0) = 0$ must be fulfilled. The solution is schematically shown in Fig. \ref{UndulatorCrystal}. It defines the $(x^\prime,z^\prime)$ coordinate system, in red color, with the $z^\prime$ axis through the turning point of the derivative, rotated clock wise by an angle $\delta$ with respect to the nominal [110] direction of the undoped diamond crystal.  Numerically $\delta$ = -0.484 mrad is obtained.

The radius of curvature is approximately given by $R_{bent} = \left(d^2x/dz^2\right)^{-1} = \pm $ 2.584 mm which turns out to be a constant for half a period, rather than of sinusoidal shape as commonly assumed in calculations of radiation emission spectra.

\begin{figure*}[tb]
\centering
    \includegraphics[angle=0,scale=0.5,clip]{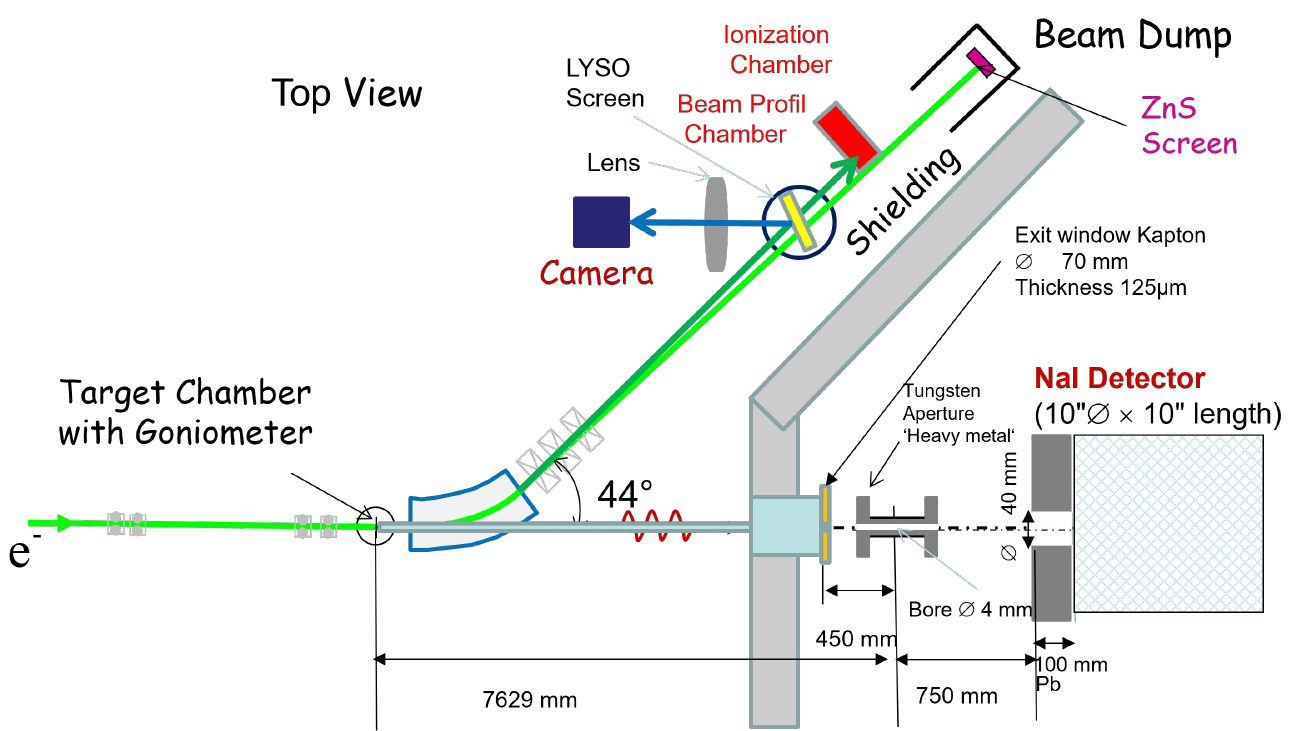}
\caption[]{Experimental setup at MAMI, not to scale. Radiation can be observed at virtually collinear geometry of the electron beam direction and the radiation detector.}\label{setup}
\end{figure*}

\section{Experimental} \label{Experiment}
The experimental setup at the Mainz Microtron MAMI is shown at Fig. \ref{setup}. The electron beam is focused onto the diamond target in the target chamber with a goniometer and thereafter deflected by a dipole magnet into the second beam profile chamber in which the scattered beam can be observed with a LYSO fluorescence screen. The target can be rotated in all 3 spatial directions with an accuracy of 35~$\mu$rad. The target is aligned using the signal of the ionization chamber that registers the scattered electrons. The orientation of the electron beam parallel to planes increases the scattering probability and, in turn, the signal of the ionization chamber. This method allows rapid alignment to a crystal plane. The radiation in the forward direction is detected with a 10 inch NaI crystal, positioned at a distance of about 8.5 m from the target. A movable cylindrical tungsten aperture with a length of 261 mm and a bore of 4~mm diameter in front of the detector defines the observation direction. An improved experiment was performed with an additional 28.8~mm long tungsten cylinder inset with an inner diameter of 2~mm,  The alignment of the aperture was done by maximizing the detected bremsstrahlung from an aluminum foil.

\begin{figure}[pos=h]
\centering
    \includegraphics[angle=0,scale=0.35,clip]{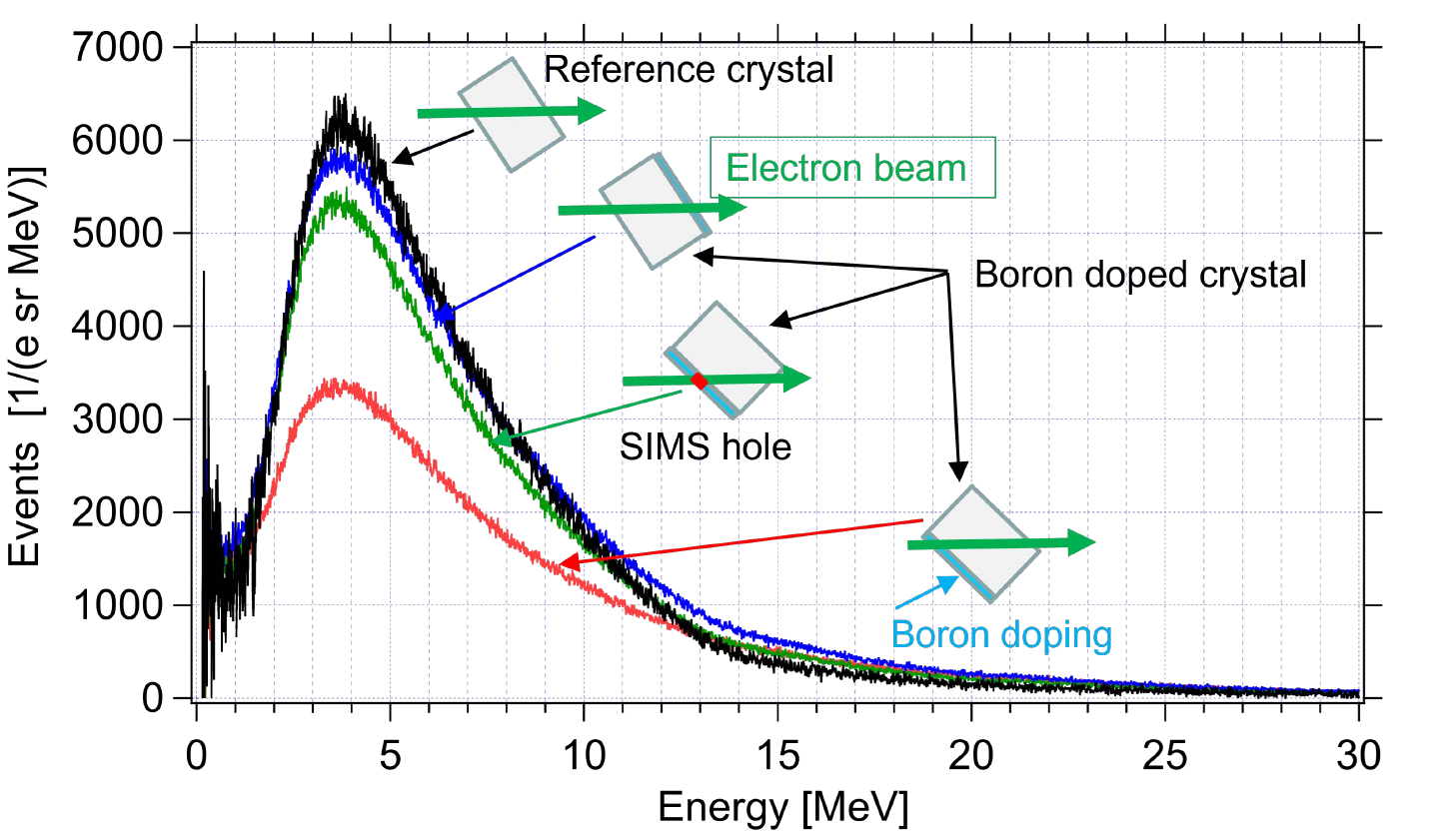}
\caption[]{Experimental radiation spectra at (110) channeling for the boron-doped diamond crystal ('Boron doped crystal'), and the reference crystal with an effective thickness of 101 $\sqrt{2}$ $\mu$m = 143 $\mu$m ('Reference crystal'). The former spectra were taken at three different geometries: (i) beam enters first the boron doped layer and thereafter the backing (red), (ii) beam enters first the backing and then the boron doped layer (blue), and (iii) beam enters the SIMS hole, at which the boron doped layer was removed (green). Spectra of random orientation of the crystal were subtracted. Spectra were taken at a typical beam current of 10~fA.} \label{spectra}
\end{figure}

Fig. \ref{spectra} depicts relevant spectra taken with the sodium iodide detector at collinear alignment of beam and tungsten aperture. The boron doped crystal and a 101~$\mu$m thick undoped reference crystal with the same surface orientation were moved into the center of the goniometer.  The boron doped layer was removed at a specific position with an area of 0.16~mm x 0.16~mm for a measurement of the doping profile with the aid of the secondary ion mass spectroscopy method (SIMS). Spectra were also taken at the position of this SIMS hole and, in addition, with the crystal rotated on an axis perpendicular to the plane of drawing by 180°. If the electron beam enters first the boron-doped layer and thereafter the straight backing crystal, the broad peak structure at about 4 MeV, which is the channeling radiation, is significantly suppressed (red spectrum). Possible explanations based on simulation calculations will be presented at the end of section \ref{simulationCalculations}.

No peak was found at this forward observation geometry near the expected energy of 1.14 MeV, as calculated with Eq. (\ref{undulatorEnergy}). There are, of course, a number of reasons for this negative result, including that with the CVD procedure the envisaged structure on the 117 $\mu$m thick host crystal was not met. However, we start with the initial working assumption that the ideal undulator structure, as described in section \ref{undulator}, has really been grown. As a guide of our considerations an simulation model, described in the next section, has been employed.

\section{The simulation model} \label{basics}
The starting point for the calculation of the radiation spectra is the following formula of J.D. Jackson's textbook "Classical Electrodynamics"
\cite[Eq. (14.65)]{Jac99}:
\begin{eqnarray} \label{Jackson14p65}
\frac{d^2 I}{d\omega d\Omega}=\frac{e^2}{4\pi^2 c}\Bigg|\int\limits_{-\infty}^\infty\frac{\hat{n}\times[(\hat{n}-\vec{\beta})\times\dot{\vec{\beta}}]}{\big(1-\vec{\beta}\cdot\hat{n}\big)^2}
e^{i\omega \big(t-\hat{n}\cdot \vec{r}(t)/c\big)} dt \Bigg|^2
\end{eqnarray}
with $d^2I$ the radiated energy with radial frequency $\omega$ into the interval $d\omega$ and solid angle element $d\Omega$, $\hat{n}$ the observation direction, $\vec{\beta}(t) = \vec{v}(t)/c$ with $\vec{v}(t)$ the particle velocity, $\vec{r}(t)$ the electron trajectory in the laboratory system, $e^2 = \hbar c \alpha$ with $\alpha$ the fine structure constant and $c$ the speed of light. For the approximate evaluation of this equation various presumptions have been made. It is assumed that (i) the electron is highly relativistic with a Lorentz factor $\gamma = 1/\sqrt{1-\beta^2}\gtrapprox$ 1000, that (ii) the deviation of the trajectory from the initial beam direction is very small, in the order of some tenfolds of Angstroms, that (iii) the undulator period $\lambda_U$ is very short, in the order of a couple of $\mu$m, that (iv) the thickness of the emitting layer is less than about 200 $\mu$m, that (v) the distance between undulator and observation point is very large, in the order of 5-10 m, that (vi) for the observation angles with respect to the beam direction $\theta\ll 1$ holds, and (vii) that the emitted photon energy is small, i.e. $\hbar\omega\ll\gamma m_e c^2$, with $m_e$ the rest mass of the electron.

Radiation emission occurs if the beam particle experiences acceleration caused by an external field, in our case it is of electrostatic origin. The acceleration components $\{a_x,a_y,a_z\}$, the prefactor of the exponential under the integral in Eq. (\ref{Jackson14p65}), are given for the above described approximations in Appendix \ref{appendix A}. It is also outlined there that the longitudinal acceleration term can be neglected, i.e. $a_z \simeq 0$. Eq. (\ref{Jackson14p65}) can be cast with the substitution $t=\zeta/c$ into the form
\begin{eqnarray} \label{Jacksonsimple}
& &
\hspace{-1.5 cm}
\frac{d^2 N \hbar\omega}{d\hbar\omega d\Omega}=\frac{\alpha}{4\pi^2}
\Bigg|\int\limits_0^{\zeta_0}
\frac{\big\{a_x(\zeta^{-1}),~a_y(\zeta^{-1}),~0\big\}}{d\zeta/dz(\zeta^{-1})}~~e^{i~\overline{\omega} ~\zeta} d\zeta \Bigg|^2
\end{eqnarray}
with
\begin{eqnarray}
& &
\hspace{-1.5 cm}
\zeta = z + \gamma^2 \int_0^{z} d z^{\prime} \big((\theta_x-\vartheta_x(z^{\prime}))^2+(\theta_y-\vartheta_y~(z^{\prime}))^2\big) \label{zeta}
\\
& &
\hspace{-0.7 cm}
d\zeta/dz=1+\gamma^2 \big( (\theta_x-\vartheta_x(z))^2+(\theta_y-\vartheta_y(z))^2\big) \label{zetainvers}
\end{eqnarray}
and
\begin{eqnarray} \label{omegabar}
\overline{\omega}=\frac{\hbar\omega}{2\gamma^2 \hbar c}.
\end{eqnarray}
In these equations $\vartheta_{x,y}(z)$ are the instantaneous angles of the electron trajectory and $\theta_{x,y}$ the observation directions, each projected onto the ($x,z$) and ($y,z$) planes. With the substitutions of Eq. (\ref{zeta}) and  (\ref{omegabar}) into the phase factor of Eq. (\ref{Jacksonsimple}) the phase integral is cast into the standard form of a Fourier representation. One consequence is that the inverse function of Eq. (\ref{zeta}), i.e. $\zeta^{-1}(z)$,  appears in Eq. (\ref{Jacksonsimple}). The integral is taken over the complete thickness $\zeta_0$ of the matter sheet which may contain an undulator and, in addition, a plane section, ensuring a coherent superposition of both parts.

The differential photon number spectrum per relative band width $d\hbar\omega/\hbar\omega$ and solid angle element $d\Omega$ is, up to a factor $\alpha/4\pi^2$, the sum of spectral power densities
\begin{eqnarray} \label{psd}
& &
\frac{d^2 N}{d\hbar\omega/\hbar\omega~d\Omega} = \frac{\alpha}{4\pi^2}\Bigg|\int_0^{\zeta_0} \frac{a_x(\zeta^{-1})}{d\zeta/dz(\zeta^{-1})} e^{i~\overline{\omega}~\zeta} d\zeta \Bigg|^2 +
\nonumber\\
& &\hspace{2.0 cm}
+\frac{\alpha}{4\pi^2}\Bigg|\int_0^{\zeta_0} \frac{a_y(\zeta^{-1})}{d\zeta/dz(\zeta^{-1})} e^{i~\overline{\omega}~\zeta} d\zeta \Bigg|^2.
\end{eqnarray}
Once the electron trajectories $\vartheta_{x,y}(z)$ were simulated and the observation angles $\theta_{x,y}$ have been chosen, the spectral power densities $|...|^2$ can be evaluated with the Mathematica 13 routine PowerSpectralDensity[data, $\omega$].


A remark seems to be appropriate why \cite[Eq. (14.65)]{Jac99} of Jackson's textbook has been used rather than the partially integrated simpler one \cite[Eq. (14.67)]{Jac99}. The reason is that phase factors occur at the analysis of the latter, resulting from non vanishing entrance and exit angles, which prevent a Fourier representation and, therefore, application of the very efficient Mathematica routine to calculate the spectral power densities.

\section{Results and Discussion} \label{calculations}
\begin{figure}[tbh]
\centering
    \includegraphics[angle=0,scale=0.65,clip]{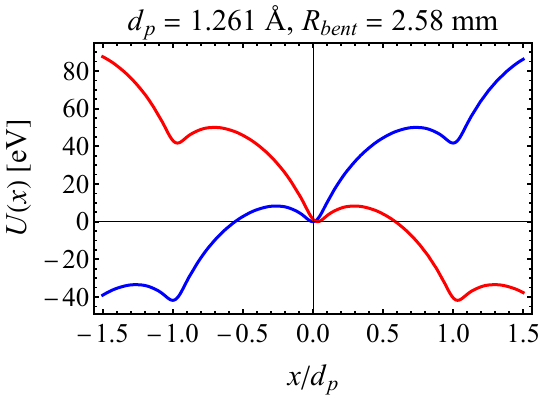}
\caption[]{Potentials for deformed (110) planes in diamond.The blue curve holds for the first half period $0\leq z\leq \lambda_U/2$ of the undulator, the red one for the second one with $\lambda_U/2< z\leq \lambda_U$. The interplanar distance is $d_p = \sqrt{2} a_C/4$ = 0.1261 nm. }\label{potentialsBentElectrons}
\end{figure}

The potential in which the particle moves was calculated as described for a plane crystal in \cite[see Fig. 2(a)]{Bac22A} to which the centrifugal potential $U_c (x)=\pm \gamma m_e c^2\beta^2/R_{bent}\cdot x$ was added. The potentials are shown in Fig. \ref{potentialsBentElectrons}. As the particle moves through the undulator crystal, the potential flips at depths $z = n \lambda/2$, with $n$ an integer, between both potential states. In the formalism described in \cite[]{Bac22A}, the scattering angles $\vartheta_x(z) = dx/dz(z)$ and $\vartheta_y(z) = dy/dz(z)$ as function of the depth $z$ must be simulated, taking into account collisions between the particle and atoms and/or electrons. These are just the quantities which are required in Eqs. (\ref{psd}) as $\vartheta_{x,y} (\zeta^{-1})$  and $\vartheta_{x,y}^{\prime} (\zeta^{-1})$ for the calculation of the radiation spectra.
\subsection{Angular Distributions} \label{calculAngDistr}
\begin{figure}[tb]
\centering
    \includegraphics[angle=0,scale=0.75,clip]{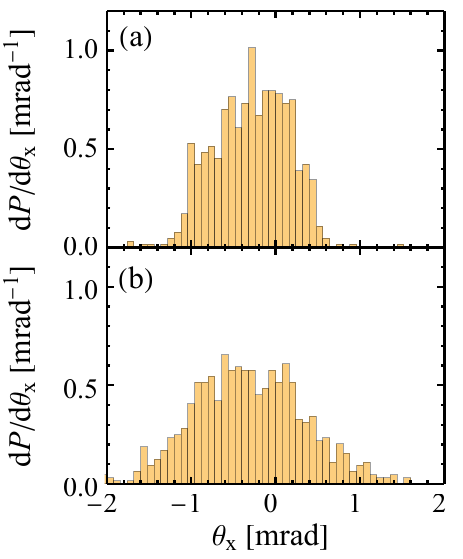}
\caption[]{Simulated scatter distributions of the electron beam, (a) only for the 4-period undulator of 20 $\mu$m thickness, (b) undulator plus 165.5 $\mu$m thick straight backing part. }\label{scatterDistributions}
\end{figure}

In the following it is assumed that the backing crystal was aligned into the beam direction, i.e., in Fig. \ref{UndulatorCrystal} at $\alpha_0$ = 45°, and that a narrow electron beam with an angular divergence $\sigma_x \simeq 0.015$ mrad and $\sigma_y \simeq 0.056$ mrad impinges the undulator crystal at vanishing small entrance angle. The scattering distribution of the electrons after passage of the undulator at $z$ = 20 $\mu$m is shown in Fig. \ref{scatterDistributions} (a). The undulator effects a dramatic broadening and also a clockwise shift of the initial electron beam distribution. With this angular distribution the electron enters the straight undoped backing crystal. As demonstrated in part (b) of the figure, the angular distribution  broadens further and apparently suffers an additional shift. A possible reason for the latter will be treated in more detail in chapter \ref{beamDeflection} below. Quantitatively, the angular distribution is rotated clockwise by about 0.4 mrad with a long extension up to -2 mrad. These facts may complicate a proper alignment of the undulator crystal into the nominal direction. It might well be that with the procedure described above the alignment results in an oblique incidence of the beam into the undulator crystal.

\subsection{Radiation spectra} \label{radiationSpectra}
Assuming that the undulator crystal was aligned properly in such a manner that the electron beam coincides with the $\alpha_0$ = 45° direction, see Fig. \ref{UndulatorCrystal}, there exist two favourable observation directions for $\theta_x$. One is the forward direction $\theta_x = 0$ in which the radiation direction defining aperture is on-axis with the electron beam direction, i.e., in Fig. \ref{UndulatorCrystal} with the $z$ axis. The other one would be the undulator radiation direction in which the aperture is rotated counter clockwise by the angle $\theta_x=\delta$ = -0.484 mrad. To make the effects clear, radiation spectra were created on the basis of the formalism described in section \ref{basics} separately for the 4-period undulator crystal alone, and for the undulator chip consisting of both, undulator part and the straight backing crystal.

\subsubsection{Model calculations} \label{modelCalculations}
\begin{figure}[tb]
\centering
    \includegraphics[angle=0,scale=0.6,clip]{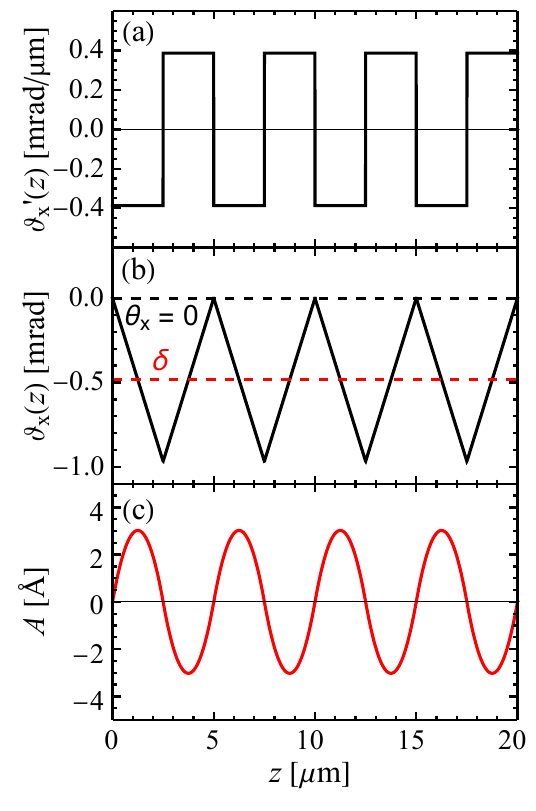}
\caption[]{Model trajectories for (a) acceleration $\vartheta_x^\prime(z)$, (b) slope of the trajectory with the initial condition $\vartheta_x(0) = \vartheta_y(0)$ = 0, and (c) amplitude $A(z)$ with $A(0)$ = 0 at observation angle $\theta_x = \delta = -0.484$  mrad, i.e. in the primed coordinate system of Fig. \ref{UndulatorCrystal}. The two favorable observation directions are included in (b) as dashed lines assigned with $\theta_x = 0$, observation in Fig. \ref{UndulatorCrystal} in $z$ direction, and $\theta_x = \delta$ at observation in $z^{\prime}$ direction. }\label{modelTrajectories}
\end{figure}

In a first step we construct an ideal model trajectory on the basis of the square wave acceleration profile neglecting any interaction with atoms and electrons. This implies that the particle sticks in the potential minimum without experiencing an excitation. In Fig. \ref{modelTrajectories} the acceleration, the derivative and the amplitude are shown, the latter with respect to the rotated primed coordinate system in Fig. \ref{UndulatorCrystal}. Indicated are as dashed lines the two favoured observation direction. The change from forward to the undulator direction is in Fig. \ref{modelTrajectories} (b) just the shift $\delta = \vartheta_x(\lambda/2)/2$ = -0.484 mrad. The radiation spectrum according to Eq. (\ref{Jacksonsimple}) can explicitly be calculated since all the quantities under the integral can be extracted from the acceleration profile $\vartheta_x^\prime(z)$.
\begin{figure}[b]
\centering
    \includegraphics[angle=0,scale=0.7,clip]{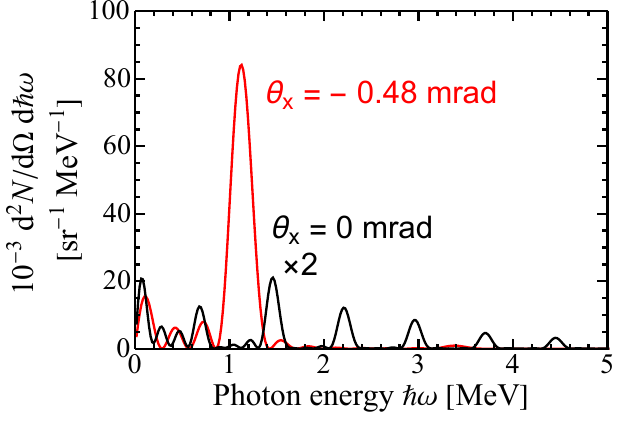}
\caption[]{Calculated photon number spectra for the 4-period undulator at $\theta_y = 0$ without radiation from the straight section of the undulator chip. Spectrum in red color for observation in undulator radiation direction, i.e., in the $z^\prime$ direction of Fig. \ref{UndulatorCrystal}, in black color observation in forward direction, i.e., in the $z$ direction of Fig. \ref{UndulatorCrystal}, magnified by a factor of two.}\label{ModelSpectra}
\end{figure}

Fig. \ref{ModelSpectra} depicts calculated photon number spectra at the two mentioned observation directions $\theta_x$. The red spectrum, for the undulator direction, i.e. the $z^\prime$ direction in Fig. \ref{UndulatorCrystal}, reveals a significant peak. The peak energy can be calculated with the well known relation
\begin{eqnarray} \label{undulatorEnergy}
\hbar\omega=k\frac{4\pi\gamma^2\hbar c}{\lambda_U \left(1+K^2/2+(\gamma\theta)^2\right)}.
\end{eqnarray}
In lowest order $k$ = 1, observation in forward direction $\theta$ = 0, amplitude $A$ = 0.302 nm,
period length $\lambda_U$ = 5.0 $\mu$m, undulator parameter $K = \gamma\cdot A\cdot 2\pi/\lambda_U$ = 0.656, a peak energy $\hbar\omega$ = 1.14 MeV is obtained in accordance with the peak energy in Fig. \ref{ModelSpectra},  red curve. The additional structures are interpreted as interferences since Eq. (\ref{Jacksonsimple}) is a phase integral. Such interferences are well known, and are described in literature for magnetic undulators with a $\big(\sin(x-x_0)/(x-x_0)\big)^2$ structure. This is an approximation for many periods, however, we are dealing with only 4 periods and such a simple formula does not hold anymore.

At observation in forward direction, $\theta_x = 0$, one naively expects from the common undulator theory a Doppler-shifted lower energy peak at 0.743 MeV with somewhat reduced intensity, and probably a second order one. However, the calculated spectrum revealed, to our surprise, complete different features, see the spectrum in Fig. \ref{ModelSpectra} in black. At a closer inspection the first and second order peaks can clearly be recognized, including a number of higher order harmonics. Indeed, such a fragmented spectrum is expected also from the magnetic undulator theory at off axis observation and larger undulator parameters $K$, see e.g. \cite[Fig.2.3]{Brau90}. It is worth noting that Eq. (\ref{Jacksonsimple}) represents all orders of the harmonics at once, i.e. it is not an expansion in terms of the harmonics.

The unexpected result at observation in forward direction, see Fig. \ref{ModelSpectra} in black, originates from the lateral displacement of the beam. As a consequence, the true reference direction of the undulator is tilted by the angle $\delta$. The reason is that the undulator was not designed with a quarter period at entrance which would rotate the main structures back into beam direction. Such a design might be somewhat more complicated in comparison with that one described in this paper. Since the results of both geometries are the same, the simpler design should be preferred. Anyway, this fact must be taken into account not only at the stage of the design of an undulator structure but, in particular, also in the course of the experiment.

\subsubsection{Simulation calculations} \label{simulationCalculations}
\begin{figure}[bt]
\centering
   \includegraphics[angle=0,scale=0.82,clip]{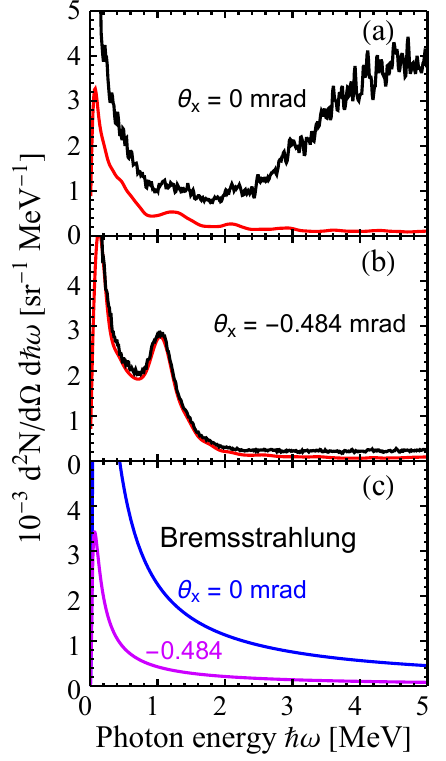}
\caption[]{Photon number spectra for the 4-period undulator including the  backing at observation in forward direction $\theta_x = 0$ mrad (a), i.e., in the $z$ direction of Fig. \ref{UndulatorCrystal}, and $\theta_x = \delta = - 0.484$ mrad (b), i.e., in the $z^\prime$ direction, both for $\theta_y = 0$. For comparison the undulator contribution alone is shown in red color. Mean of 400 sample trajectories. Panel (c) depicts the bremsstrahlung contribution \cite[Eq. 10.2]{AkhS98} for an isotropic distribution of the carbon atoms at two observation angles $\theta_x$ = 0 and -0.484 mrad with respect to the beam direction.} \label{NumberSpectra}
\end{figure}

In this subsection we describe full-fledged simulation calculations in the real potential taking into account interactions of the beam electron with atoms and bulk electrons. In Fig. \ref{NumberSpectra} the results are shown. The middle part (b) shows spectra at observation in undulator direction, i.e. at $\theta_x = \delta = - 0.484$ mrad. The undulator peak remains clearly visible, however, it is a factor of about 50 weaker. Notice, the scale has been magnified by a factor of 20 in comparison with Fig. \ref{ModelSpectra}. The upper part (a) shows spectra at observation in forward direction $\theta_x = 0$. There seem to be indications of broad structures at positions of the first and second harmonic, and perhaps also higher ones as indicated by the spectrum in red color. The weak intensity suggests strong de-channeling originating from rather shallow (110) potential as shown in Fig. \ref{potentialsBentElectrons}. The latter is a consequence of too strong a boron doping of the undulator, assuming a lattice expansion as taken from Fig. \ref{BoronCharacteristics}.

At observation in undulator direction the channeling radiation from the rather thick backing of 165.5 $\mu$m is strongly suppressed. However, it must be mentioned that the calculations have been performed for ideal experimental conditions. In a real experiment the dimension of the detector aperture in the order of 0.5 mrad results in an increase of the channeling radiation background. In addition, the low energy noise may be larger, and also tails of the channeling radiation originating from the detector response function impairs these predictions. Finally, also the bremsstrahlung background deteriorates the peak-to-background ratio. It is shown in Fig. \ref{NumberSpectra} (c) for a straight electron trajectory at two different observation angles. The contributing background may be expected somewhere in between the limiting spectra for 0 and -0.484 mrad. For the latter observation angle it is also suppressed.
\begin{figure}[t]
\centering
   \includegraphics[angle=0,scale=0.75,clip]{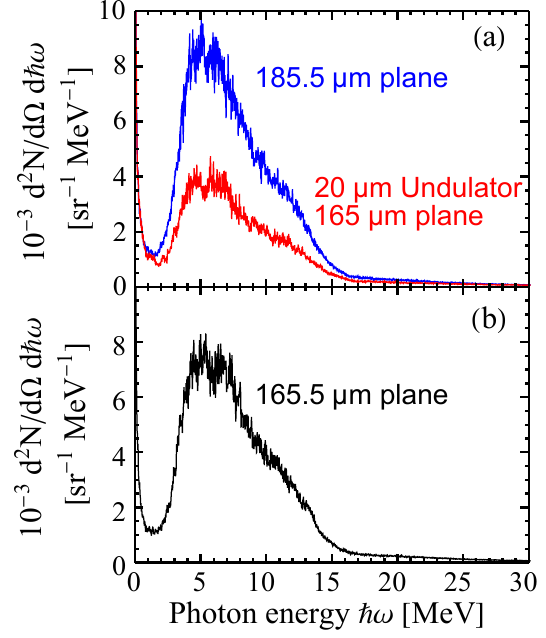}
\caption[]{(a) Photon number spectra for an effective 185.5 $\mu$m thick plane crystal in (110) orientation at observation in forward direction, i.e. at $\theta_x=\theta_y = 0$ mrad (blue), together with the spectrum for a 4-period 20 $\mu$m thick undulator crystal combined with an 165.5 $\mu$m thick plane backing crystal behind (red). (b) Spectrum of a 165.5 $\mu$m thick plane crystal with a broadened entrance distribution of the beam of $\sigma_x$ = 0.137 mrad which was assumed to simulate a 20 $\mu$m amorphous entrance layer. Mean of 400 sample trajectories.} \label{NumberSpectra30MeV}
\end{figure}

\begin{figure}[t]
\centering
   \includegraphics[angle=0,scale=0.33,clip]{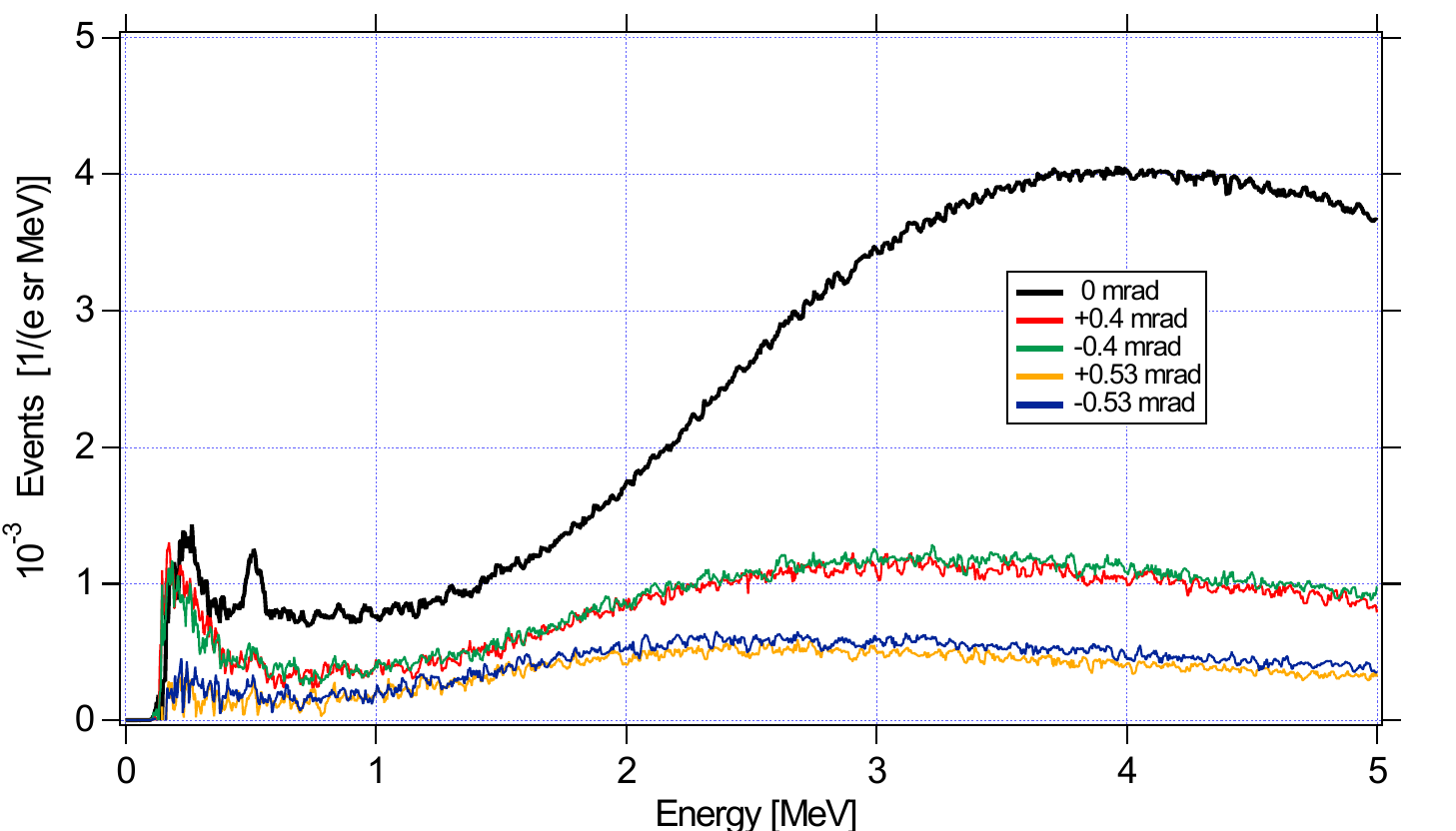}
\caption[]{Peak search with the experimental setup shown in Fig. \ref{setup}, however, with a decreased aperture diameter of 2 mm. The spectra are corrected for a background contribution taken by de-tuning the angle far away from the channeling condition. Shown are spectra for the forward direction, 0 mrad (black), and a de-tuning by $\theta_x = \pm$ 0.4 mrad and $\pm$ 0.53 mrad, as assigned by the color code. It is worth mentioning that the peak shout appear at angles with negative sign only, since at positive sign the undulator is de-tuned into the background region. Peak at 0.511 MeV is due to positron annihilation.} \label{ExpPeakSearch}
\end{figure}

The experimental spectra depicted in Fig. \ref{spectra} show a significant difference if the beam enters first into the undulator crystal and propagates thereafter into the flat backing one (red), in comparison with the reversed orientation of the chip (blue). Such an effect is expected from the broadening and shift of the scattering distribution originating from the undulator crystal as shown in Fig. \ref{scatterDistributions} (a). This distribution is of significant difference in comparison with the narrow beam distribution if entering the backing crystal at the reversed configuration. Simulation calculations for both cases are shown in Fig. \ref{NumberSpectra30MeV}. They reproduce quite well the intensity ratio suggesting the existence of curved structures which deflect the beam. The simulated spectrum for a complete amorphous layer at the first 20 $\mu$m, shown in Fig. \ref{NumberSpectra30MeV} (b), is at variance with the observation, see Fig. \ref{spectra}, red spectrum, and can be excluded. This fact prompted us to search for the peak structure predicted in Fig. \ref{NumberSpectra} (b) with improved experimental conditions. The aperture diameter was reduced from 4 to 2 mm diameter and the statistics significantly enhanced. The result is shown in Fig. \ref{ExpPeakSearch}.

Our method is rather sensitive since the peak should appear only at negative observation angles while the positive ones signify background. No indication of a peak or even a tiny enhancement around 1.1 MeV was observed. From a statistical point of view a few percent of the structure shown in Fig. \ref{NumberSpectra} (b) should have been appeared. The reason for this result might be that the peak is in reality much weaker and broader than depicted in case, e.g., of a rapid relaxation of the undulating amplitude, with the consequence that it finally does not pops out from the unavoidable background radiation. Alternative explanations may be that the assumed lattice expansion as function of the boron concentration might not be correct, or that randomly oriented microstructures may have been grown during CVD grown which broaden the angular distribution much more than an amorphous layer does. It can also not be excluded that our simulation model somewhat overestimates the peak structure shown in Fig. \ref{NumberSpectra} (b).

We mention that the simulation calculations for the flat crystal, Fig. \ref{NumberSpectra30MeV} (b)  with a thickness of 165.5 $\mu$m, does not perfectly reproduce the measured channeling radiation distribution. In particular, the simulated spectrum is shifted to higher energy, and the absolute peak intensity is about 20 \% larger. It is unlikely that the reason can be found in the experimental response function of the $10^{\prime\prime}$ NaI detector, although a de-convolution could shift the spectrum somewhat to higher energies. Instead, we attribute this fact mainly to deficiencies of our model assumptions. However, since the model was developed only for the purpose to guide us during the course of an experiment online, such a variance results not in serious misdirections.

\begin{figure}[b]
\centering
    \includegraphics[angle=0,scale=0.2,clip]{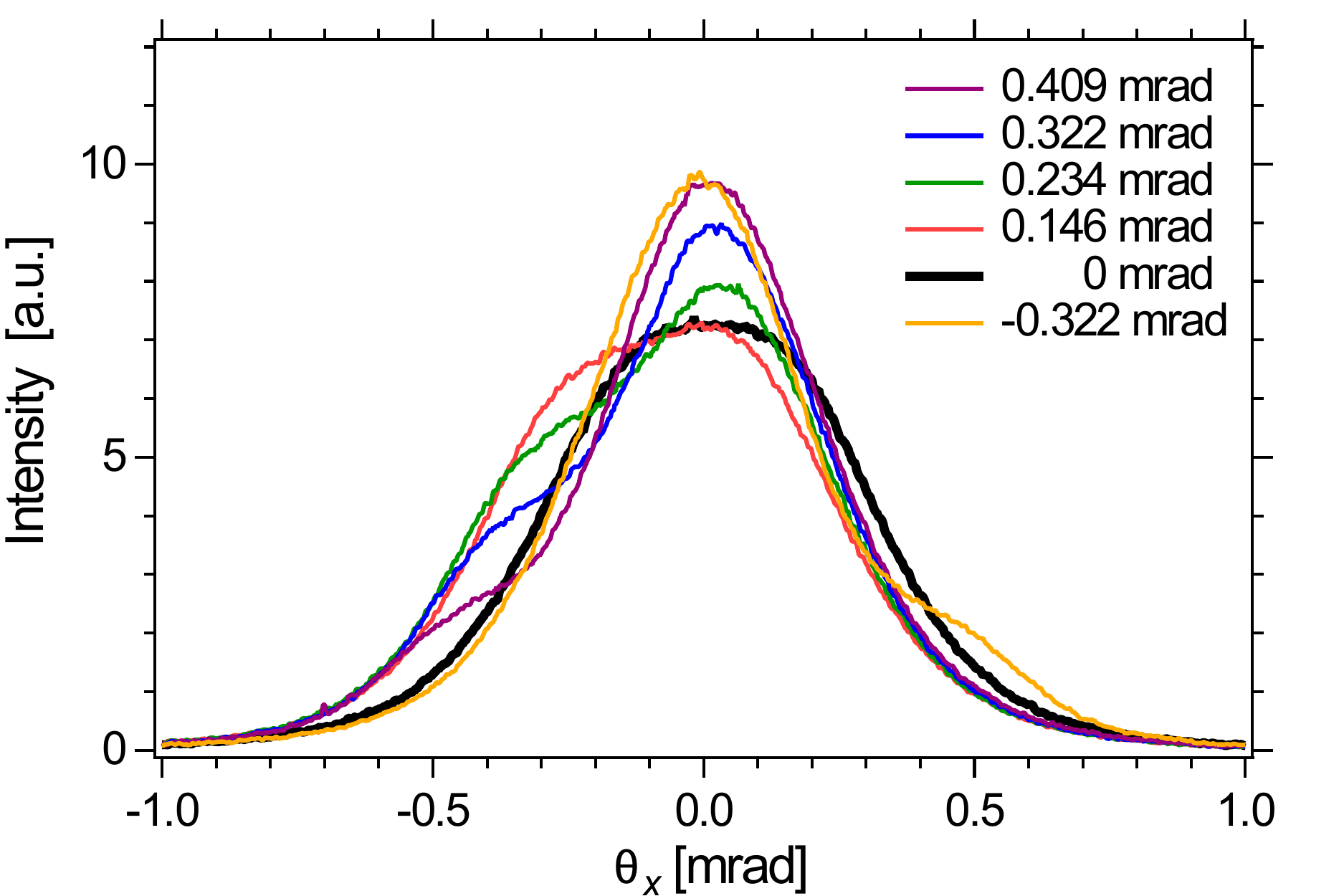}
\caption[]{Projected experimental scatter distributions for the 855 MeV electron beam impinging on a (50$\pm 2)~\mu$m thick plane diamond crystal, effective thickness for the (110) plane (70.7$\pm 2.8)~\mu$m, at tilt angles $\psi_x$ as indicated by the color code. The spatial horizontal and vertical $x, y$ coordinates were calibrated with a special target with known position markers, and converted into angles with effective lengthes from  target to the screen of 4.93 and 6.02 m, respectively.}  \label{beamDeflectionExperiment}
\end{figure}
\section{Asymmetric scatter distributions at oblique incidence on the (110) plane of diamond} \label{beamDeflection}
Downstream the target the scatter distribution can be observed with a LYSO fluorescence screen and a camera system as function of the horizontal and vertical angles ($\theta_x,\theta_y$), respectively, see Fig. \ref{setup}. The scattered electron impinges the screen with dimensions 28 mm $\times$ 13 mm $\times$ 0.1 mm perpendicular to its surface. The fluorescence light is observed horizontally at an angle of 22.5° with a zoom optics which focuses the light onto a 8.45 mm $\times$ 7.07 mm sensor with 2,448 $\times$ 2,048 pixels, positioned at an effective distance of about 762 mm. For a (50$\pm 2)~\mu$m thick diamond single crystal, corresponding to an effective thickness in [110] direction of (70.7$\pm 2.8)~\mu$m, the projected scatter distribution $dP/d\theta_x$ was measured as function of the incidence angle $\psi_x$ between electron beam and the (110) plane. The latter corresponds to a rotation around the $y$ axis, see Fig. \ref{UndulatorCrystal}. To avoid axial channeling at strings in the (110) plane, the crystal was rotated around the $x$ axis, see Fig. \ref{UndulatorCrystal}, by an angle $\psi_y$ between -15° and -11.5°, i.e. at a large complementary angle with respect to the [001] direction. As depicted in Fig. \ref{beamDeflectionExperiment} an asymmetric scatter distribution is observed.

\begin{figure}[t]
\centering
    \includegraphics[angle=0,scale=0.4,clip]{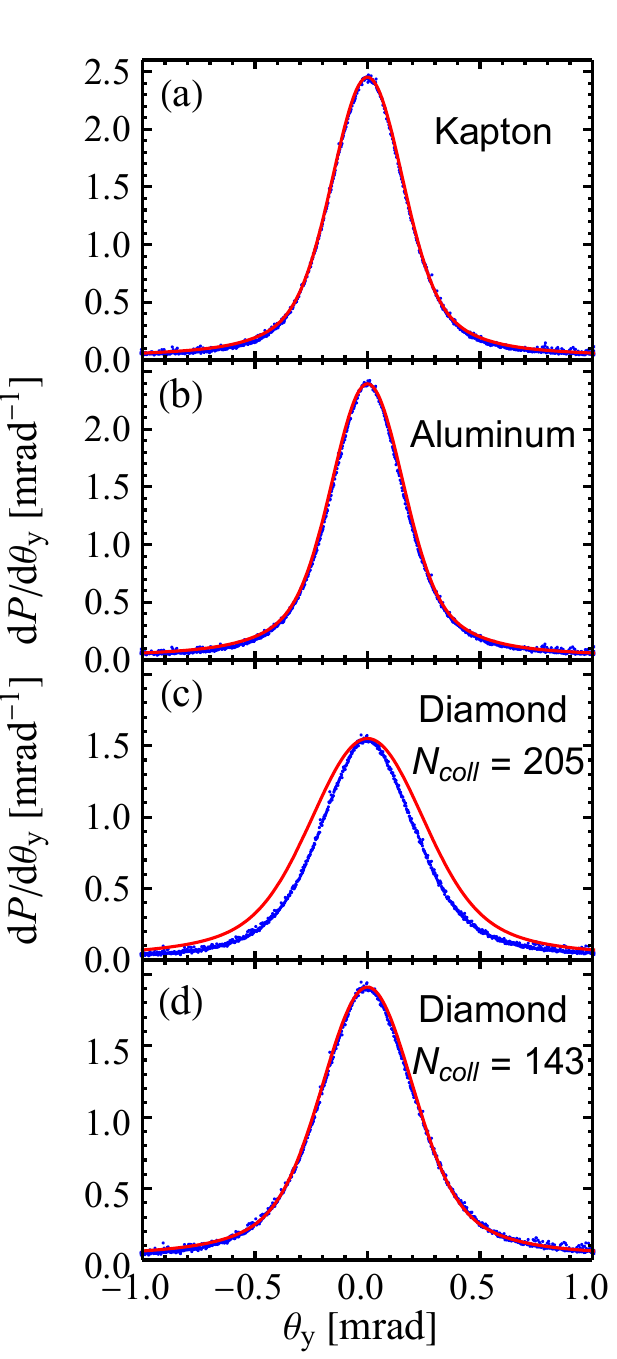}
\caption[]{Projected scatter distributions for the 855 MeV electron beam impinging on a 75 $\mu$m Kapton foil (a), a 25 $\mu$m aluminum foil (b), a (50$\pm 2)~\mu$m thick plane diamond crystal, corresponding to an effective thickness for the (110) plane of (70.7$\pm 2.8)~\mu$m (c) and (d). Blue points represent experimental data, red full lines calculations with the formalism of Molière \cite{Mol48} with collision numbers $N_{coll}$ = 204.8 (c) and $N_{coll}$ = 143.4 (d). The calculated spectra were convoluted with the point spread function of the LYSO fluorescence screen to make a comparison with experimental data meaningful.} \label{scatterDistKaptonAlDiamond}
\end{figure}
\begin{figure}[t]
\centering
    \includegraphics[angle=0,scale=0.52,clip]{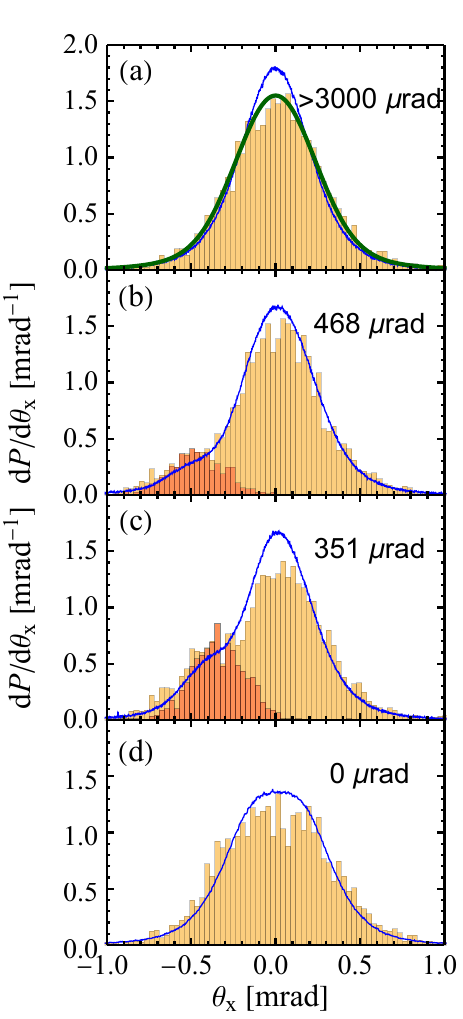}
     \includegraphics[angle=0,scale=0.52,clip]{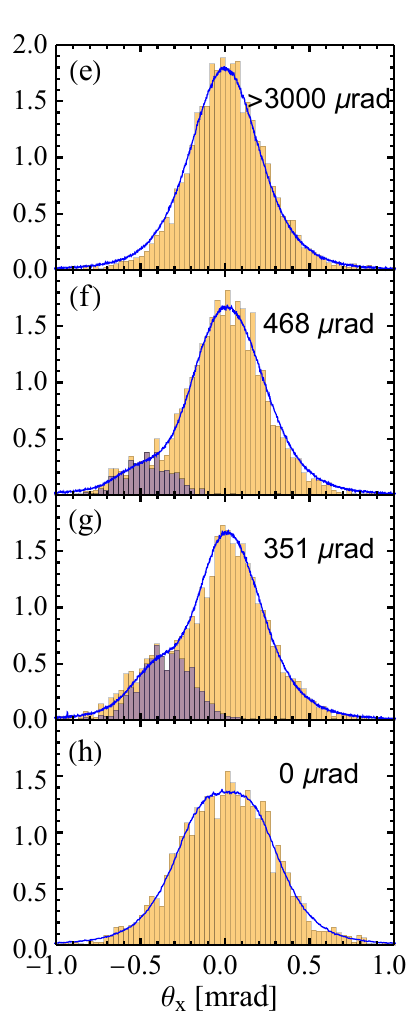}
\caption[]{Scatter distributions of an 855 MeV electron beam impinging on a 50 $\mu$m thick plane diamond crystal, effective thickness for the [110] direction  plane 70.7 $\mu$m with tilt angles $\psi_x$ as assigned in the sub pictures. The histograms are simulation calculations, (a)-(d) with a projected scattering distribution for amorphous matter, (e)-(h) for a reduced mean free path length by a factor of 1/0.7 = 1.43. The brown curve in (a) is a calculation of the projected scatter distribution according to Molière \cite{Mol48}. The red-brown overlays in (b) and (c) as well as the blue ones in (f) and (g) are the fractions of particles which experienced channeling or temporarily channeling in the last 14.7 $\mu$m of the crystal. Simulations with 3000 samples. The blue curves are experimental data taken with a LYSO fluorescence screen with dimensions 34~mm $\times$ 36~mm $\times$ 0.2 ~mm, not de-concoluted from the point spread function.} \label{beamDeflectionSimulation}
\end{figure}

At Fig. \ref{scatterDistKaptonAlDiamond} (c) the measured scatter distribution at such a random impact is compared with theoretical predictions of Molière for the projected distribution \cite{Mol48}. (We have included the factor $Z(Z+1)$ instead of $Z^2$ in the quantity $\chi_c^2$ as suggested by Bethe \cite{Bet53} to account for scattering at electrons. This factor has also been adapted by Lynch and Dahl \cite{LynD91}.) The width of the measured distribution depicted in blue color is 21 \% smaller. To check correctness of our calculation, Fig. \ref{scatterDistKaptonAlDiamond} (a) and (b) show comparisons with effective amorphous materials like Kapton and aluminum, respectively. Very good agreement between Molière's theory \cite{Mol48} and experiment is found. This observations let us conclude that the reduction of the line width originates from the crystalline structure of diamond, despite the random impact of the beam onto the crystal.

Taking advantage of the scheme to calculate cross-sections in crystals, see the text book of Ter-Mikaelian \cite[in paricular pages 34-67]{TerM85} and also \cite{MazS20}, the scattering cross-section may be written as
\begin{eqnarray}
& &
\hspace{-0.7 cm}
\frac{d\sigma}{d\Omega} = N \frac{d\sigma_{am}}{d\Omega}\Big(1-\exp\Big[-\frac{q^2 \overline{u^2}}{\hbar^2}\Big]\Big)+\frac{d\sigma_{int}}{d\Omega}, \label{cross-section}
\\
& &
\hspace{-1.2 cm}
\frac{d\sigma_{int}}{d\Omega} = \frac{d\sigma_{am}}{d\Omega}\exp\Big[-\frac{q^2 \overline{u^2}}{\hbar^2}\Big]\Bigg| \sum_i \exp\Big(\imath~\frac{\overrightarrow{q} \cdot\overrightarrow {r}_{i0}}{\hbar} \Big) \Bigg|^2  \label{cross-sectionInterference}
\end{eqnarray}
with $d\sigma_{am}/d\Omega$ the cross-section for scattering in the amorphous medium. In case of a random orientation the sum of the random phases from the atoms at positions $\overrightarrow {r}_{i0}$ interferes to zero, and the cross-section reduces by the factor in round parentheses of Eq. (\ref{cross-section}). It is a function of the mean square of the thermal vibrations $\overline{u^2}$ and the square of the momentum transfer $\overline{q^2}$. This fact may be interpreted as a reduction of the mean number of collisions $N_{coll}$. Indeed, as shown in Fig. \ref{scatterDistKaptonAlDiamond} (d), the measured scatter distribution can well be described also with Molière's theory with a collision number reduced  by a factor of 0.7. In case that the kinematical conditions approach channeling, the interference term, Eq. (\ref{cross-sectionInterference}), can not be neglected, resulting in the side peaks shown in Fig. \ref{beamDeflectionExperiment}. The corresponding theoretical calculation is beyond the scope of this work. Instead, we have developed a simple model to describe them.

The model is based on the observation that the side peaks seem to be correlated to the entrance tilt angles $\psi_x$. This conjecture suggests that the beam deflection might be interconnected with the channeling phenomenon. For instance, at $\psi_x$ = 0.351 mrad the initial transverse energy amounts to 53 eV meaning that the electron moves in a way of above barrier channeling over the (110) potential walls with a depth of 24 eV. At scattering close to the potential minimum, where the atomic density is largest, the electron may be captured and continues propagation in a real channeling mode. In that case it may exit the crystal at about the entrance tilt angle with a broadened distribution depending on the transverse energy and the phase of the oscillation at exit. In case a re-channeled particle de-channels again, some memory effects may remain that it was in between in the channeling mode. Such a conjecture is in accordance with simulation calculations selecting only events which were temporarily in the channeling mode. As demonstrated in Fig. \ref{beamDeflectionSimulation} (b) and (c), the mean angle of the red-brown overlay distribution is clearly correlated with the entrance tilt angle. The extra events  at the largest negative scattering angles probably originate from an earlier re-channeling - de-channeling process.

It is worth mentioning that our simulation distribution as shown in Fig. \ref{beamDeflectionSimulation} (a) is in very good agreement with the calculation on the basis of Molière's theory which is shown by the brown full curve.


In Fig. \ref{beamDeflectionSimulation} (e)-(h) results of simulation calculations are depicted which were performed with a factor of 1/0.7 = 1.43 increased mean free path length between two atomic scattering events. Such a heuristic approach describes nearly perfectly the measurements for the selected tilt angles $\psi_x$, however, are not in such a good agreement for smaller ones, not shown here. Though quite different kinematical conditions were chosen than those in A. Mazzolari and A. Sytov  et al. for silicon \cite[Fig. 4 c]{MazS20}, the general features apparently are  preserved for the (110) plane. Anyway, a rigorous treatment of the effect is beyond the scope of this work.

\section{Conclusions} \label{Conclusions}
A boron-doped 4-period diamond undulator with a thickness of 20 $\mu$m, grown on a plane (110) crystal with an effective thickness of 165.5 $\mu$m, was investigated experimentally with the aid of the 855 MeV electron beam of MAMI. The observation geometry for the search of an undulator peak was optimized with the aid of simulation calculations. The latter are based on the continuum potential picture and an equation taken from Jackson's textbook of classical electrodynamics which explicitly evaluates the acceleration of the simulated particle trajectory. An undulator peak was not observed. A number of reasons for this negative result were pursued, including the too shallow (110) potential shown in Fig. \ref{potentialsBentElectrons}, and  that various design parameters for the structure of the real undulator were not met. In conclusion, even experiments with the high quality 855 MeV electron beam of MAMI, supported by simulation calculations for the choice of optimal observation parameters, resulted only in rather vague conjectures about the prepared undulator structure.

\begin{figure}[pos=h]
\centering
    \includegraphics[angle=0,scale=0.57,clip]{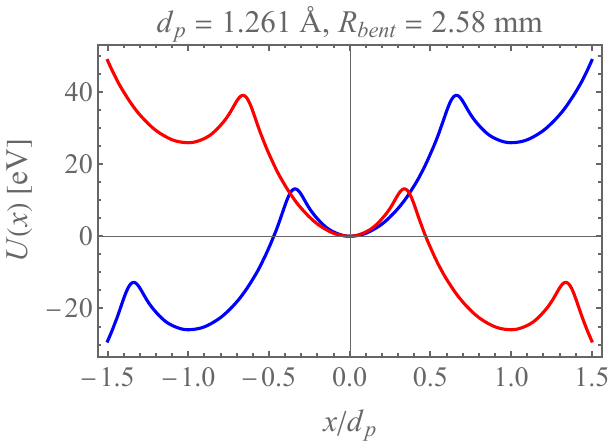}
\caption[]{Potentials for deformed (110) planes in diamond for positrons. The blue curve holds for the first half period $0\leq z\leq \lambda_U/2$ of the undulator, the red one for the second one with $\lambda_U/2< z\leq \lambda_U$. The interplanar distance is $d_p = \sqrt{2} a_C/4$ = 0.1261 nm. }\label{potentialsBentPositrons}
\end{figure}

\begin{figure}[pos=h]
\centering
    \includegraphics[angle=0,scale=0.70,clip]{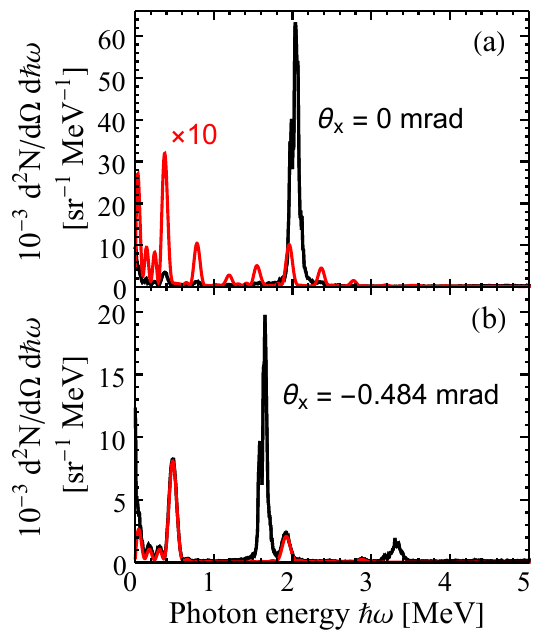}
\caption[]{Same as Fig. \ref{NumberSpectra} (a) and (b) for 530 MeV positrons. Simulated photon number spectra are shown for forward direction (a), and at $\theta_x = \delta = - 0.484$ mrad at $\theta_y = 0$ (b). Black curves for the undulator chip including the backing crystal with an effective thickness of 165.5 $\mu$m. For comparison the contribution of the 4-period undulator with a period length of 5 $\mu$m is shown in red color. Mean of 100 sample trajectories. } \label{numberSpectrumPositrons}
\end{figure}

Deeper insight into the features of the boron-doped diamond undulator may be gained with positrons. A 530 MeV beam is now available at MAMI which was constructed on the basis of \cite{BacL22}. In comparison with electrons the potential is for a large fraction of the impact coordinate nearly harmonic, see Fig. \ref{potentialsBentPositrons}, and simulation calculations reveal rather narrow line structures as shown in Fig. \ref{numberSpectrumPositrons}. The photon energy of the undulator peak amounts due to the reduced beam energy to only 0.48 MeV, cf. Fig. \ref{numberSpectrumPositrons} (b). The energy integrated intensity in the peak is about $10^3$/sr and positron.

Scatter distributions were measured for a 75 $\mu$m Kapton, a 25 $\mu$m aluminum foils, and a 70.7 $\mu$m diamond plate in random orientation. The results were compared with Molière's scatter theory for amorphous medii. Very good agreement was found for Kapton and aluminum while for diamond the experimental width is reduced by 21\%. This effect is interpreted as coherent scattering suppression in single crystals.

The scatter distribution at tilted injection of the beam into the (110) plane of a 50$\cdot\sqrt{2}$ $\mu$m thick plane diamond crystal shows a clear asymmetry which resembles partial beam deflection. This phenomenon originates probably also from coherence effects. However, simulation calculations suggest that the effect may also be interpreted by a re-channeling of the incident particle. On the basis of this assumption a simple heuristic model was developed which describes the experimental scatter distributions quite well.

In conclusion, calculations as those presented in this paper, which are performed all with the aid of a personal computer, may be an on-line guide during the course of future experiments.

\section*{Acknowledgements} \label{Acknowledgements}

Fruitful discussions with A. V. Korol, A. V. Solov'yov, and A. Sytov are gratefully acknowledged.

\section*{Declarations}
This work has been financially supported by the European Innovation Council (EIC) Pathfinder TECHNO-CLS project 101046458.

\section*{Data availability statement}
All data are freely available.

\section*{Appendix}
\begin{appendix}
\section{Spectral power density in terms of angular variables} \label{appendix A}
In this appendix some details are presented how the spectral power density of Eq. (\ref{Jacksonsimple}) comes about and which accuracy may be expected in applying it. This expression is an approximation of Eq. (\ref{Jackson14p65}) for which according to the restrictions made in the text the following power series expansions for the three vector components assigned as $\{a_x,~a_y,~a_c\}$ were used
\begin{eqnarray} \label{nhat}
& &\hspace{-0 cm}
\hat{n}=\{\theta_x,~\theta_x, ~1 - \theta_x^2/2- \theta_y^2/2\},
\nonumber\\
& &\hspace{-0 cm}
\vec{\beta}=\beta \big\{\vartheta_x,~ \vartheta_y, ~1 - \vartheta_x^2/2- \vartheta_y^2/2\big\},
\nonumber\\
& &\hspace{-0 cm}
\dot{\vec{\beta}}=\beta~c \{\vartheta_x', ~\vartheta_y',~ -(\vartheta_x \vartheta_x'+\vartheta_y \vartheta_y')\},
\nonumber\\
& &\hspace{-0 cm}
\beta=1 -\frac{1}{2\gamma^2},
\end{eqnarray}
The angles $\theta_{x,y}$ signify the projected observation directions onto the ($x,z$) and ($y,z$) planes, respectively, and $\vartheta_{x,y}^{\prime}(z)$ the derivatives of $\vartheta_{x,y}(z)$ with respect to the z coordinate of the electron trajectory. The result is
\begin{eqnarray} \label{equationA1}
& &\hspace{-0 cm}
\frac{\hat{n}\times[(\hat{n}-\vec{\beta})\times\dot{\vec{\beta}}]}{\big(1-\vec{\beta}\cdot\hat{n}\big)^2}=\{a_x,a_y,a_z\}=
\nonumber\\
& &\hspace{-1.5 cm}
= 2 \gamma^2 \Big\{\frac{-1 + \gamma^2 \big((\theta_x-\vartheta_x)^2 -(\theta_y-\vartheta_y)^2\big)}{\big[1 + \gamma^2 \big((\theta_x-\vartheta_x)^2 +(\theta_y-\vartheta_y)^2\big)\big]^2}\cdot c \vartheta_x'+
\nonumber\\
& &\hspace{0.5 cm}
+\frac{2\gamma^2 (\theta_x-\vartheta_x)~(\theta_y-\vartheta_y)}{\big[1 + \gamma^2 \big((\theta_x-\vartheta_x)^2 +(\theta_y-\vartheta_y)^2\big)\big]^2}\cdot c \vartheta_y',
\nonumber\\
& &\hspace{-1 cm}
\frac{2\gamma^2 (\theta_x-\vartheta_x)~(\theta_y-\vartheta_y)}{\big[1 + \gamma^2 \big((\theta_x-\vartheta_x)^2 +(\theta_y-\vartheta_y)^2\big)\big]^2}\cdot c \vartheta_x'+
\nonumber\\
& &\hspace{0.5 cm}
+\frac{-1 - \gamma^2 \big((\theta_x-\vartheta_x)^2 -(\theta_y-\vartheta_y)^2\big)}{\big[1 + \gamma^2 \big((\theta_x-\vartheta_x)^2 +(\theta_y-\vartheta_y)^2\big)\big]^2}\cdot c \vartheta_y',
\nonumber\\
& &\hspace{-0.5 cm}
~\mathcal{O}(\theta_x,\theta_y,...)\cdot c\vartheta_x'+\mathcal{O}(\theta_x,\theta_y,...)\cdot c\vartheta_y' \Big\}.
\end{eqnarray}
In this approximation second order terms and higher order ones in $\theta_x,\theta_y, \vartheta_x, \vartheta_y$ as well as cross terms were neglected since their contributions are very small in comparison with counterparts multiplied by $\gamma$. The $z$ component is symbolically introduced by $\mathcal{O}(\theta_x,\theta_y,...)$. It vanishes at on-axis observation and is otherwise of first order either for terms in $\theta_{x,y}$ and of third order in angular coordinates for terms of order $\gamma^2$, and is therefore negligible small in comparison with the $x,y$ components.
In addition, the longitudinal acceleration originating from the momentum coupling between the transverse and longitudinal degrees of freedom, resulting in the well known 'figure eight' motion, has been neglected with the same reasoning as given above taking into account that $p_z = \beta\gamma m_e c(1-\vartheta_x^2/2)$.

Special care is required to exploit the exponential expression $\big(t-\hat{n}\cdot \vec{r}(t)/c\big)$. The position vector is written as
\begin{eqnarray} \label{rpos}
\overrightarrow{r}(t) = \beta c \int_0^t dt' \big\{ \sin\vartheta_x(t') , \sin\vartheta_y(t'),\cos\vartheta(t')\big\}
\end{eqnarray}
resulting in
\begin{eqnarray} \label{equationA2}
\big(t-\hat{n}\cdot \vec{r}(t)/c\big) = \int_0^t dt'\big(1-\beta\sin\theta_x\sin\vartheta_x(t')-
\nonumber\\
& &\hspace{-6.0 cm}
-\beta\sin\theta_y\sin\vartheta_y(t')-\beta\cos\theta\cos\vartheta(t')\big)
\nonumber\\
& &\hspace{-8.4 cm}
\cong \frac{1}{2\gamma^2}\Big( t+\gamma^2\int_0^t dt'\big((\theta_x-\vartheta_x(t'))^2+(\theta_y-\vartheta_y(t'))^2 \big)\Big).
\end{eqnarray}
The expansion has been carried out with $\theta^2=\theta_x^2+\theta_y^2$ and $\vartheta^2=\vartheta_x^2+\vartheta_y^2$ up to the second order as above. After the substitution $t=\zeta/c$ the expressions Eq. (\ref{zeta}) in combination with Eq. (\ref{omegabar}) is obtained. Also this expansion is rather accurate since higher order terms contribute in forth order.

It should be mentioned that in \cite{Bac24} the spectral power density was calculated with an incorrect equation, rendering Fig. 14 and 18 to be obsolete.
\end{appendix}




%
\bio{}
\bibliographystyle{elsarticle-num}
\bibliography{bibfileBa}
\endbio

%
%

\end{document}